# Relativistic addition of parallel velocities from time dilation

Bernhard Rothenstein, "Politehnica" University of Timisoara, Physics Department, Timisoara, Romania

Stefan Popescu, Siemens, Erlangen, Germany

*Abstract. The relativistic addition of parallel velocities is derived involving relativity only via the time dilation formula, avoiding the length contraction used by many authors in conjunction with time dilation. The followed scenario involves a machine gun that fires successive bullets, considered from its rest frame and from the rest frame of the target, the bullets hit.*

Derivations of the formula for the relativistic addition of parallel velocities (RAPV) without using the Lorentz transformations are known. Some of them[1,2,3] derive the RAPV using both Lorentz contraction and time dilation. Peres[4] presents a derivation of the RAPV based on the relativistic Doppler formula using light signals as detected by three inertial observers, one of which is at rest relative to the light source that emits the successive light signals at constant time intervals, while the two other observers move with speeds $u_x$ and $u'_x$ relative to it, respectively. Sartori[5] presents a derivation of the RAPV involving relativity via the time dilation. The scenario he follows presents many side steps and complicated figures.

We propose a scenario that leads to RAPV involving only the time dilation formula which, in Asher's[4] derivation, is a direct consequence of the relativistic principle only. This scenario involves the OX(O'X') axes of the K(XOY) and K'(X'O'Y") inertial reference frames in the standard arrangement, with the corresponding synchronized C(x,0) and C'(x',0) of the two frames located at the different points of the mentioned axes. An important role is played by the clocks $C_0(0,0)$ and $C_{0'}$ located at the origins O and O', respectively. As a consequence of the time dilation effect, when clock $C_{0'}(0,0)$, which reads t', is located in front of clock C(x,0), which reads t, the two clock readings are related by

$$t = \frac{t'}{\sqrt{1-V^2/c^2}} = \gamma(V)t' \qquad (1)$$

where V represents the speed of K′ relative to K. For the same reasons, when clock $C_0(0,0)$, reading t, is located in front of a clock C′(x′,0), reading t′, the two clock readings are related by

$$t' = \frac{t}{\sqrt{1-V^2/c^2}} = \gamma(V)t. \tag{2}$$

The scenario also involves a machine gun (MG) located at the origin O and at rest relative to K. It fires successive bullets in the positive direction of the OX axis at constant time intervals $T_e$. Let $u_x$ be the speed of the bullets relative to K (Figure 1). A target at rest located at the origin of K′, O′, receives the first fired bullet at a time t=0, when it is located in front of the MG. The second bullet, having been fired at a time $T_e$, hits the target at a point M(x=$VT_r$,0), where $T_r$ represents the reading of clock C(x=$VT_r$,0) located at that point. Equating the distance travelled by the second bullet with the distance travelled by the target between the reception of the first and the second bullets, we obtain

$$u_x(T_r - T_e) = VT_r \tag{3}$$

from where we obtain

$$T_r/T_e = (1 - V/u_x)^{-1}. \tag{4}$$

At that very moment, clock $C_{0'}(0,0)$ which is commoving with the target, reads $T'_r$, which is related to $T_r$ by (1)

$$T_r/T'_r = \gamma(V). \tag{5}$$

Combining (4) and (5) we obtain

$$T'_r/T_e = \gamma^{-1}(1 - V/u_x)^{-1}. \tag{6}$$

Equation (6) accounts for the Doppler Effect because it relates two proper time ($T_e$-0) and ($T_r$′-0).

Detecting the same experiment from the rest frame of the target (K′), the MG moves with speed V in the negative direction of the O′X′ axis while the bullets move with speed $u_x$′ in the positive direction of the same axis. Equating the distance travelled by the MG between the firing of two successive bullets with the distance travelled by the second fired bullet, we obtain

$$u'_x T'_e = V(T'_r - T'_e) \tag{7}$$

where $T_e$′ and $T_r$′ represent the readings of the clocks $C_{0'}(0,0)$ and C′(x′,0) respectively. From (7) we obtain

$$T'_e/T'_r = (1 + V/u'_x)^{-1}. \tag{8}$$

When clock $C_0(0,0)$, which is commoving with the MG, reads $T_e$ it is located in front of a clock $C'(x',0)$ reading $T_e'$; The two clock readings are related by (2)

$$T_e' = \gamma(V) T_e \tag{9}$$

with which (8) becomes

$$T_e / T_r' = \gamma^{-1}(1 + V/u_x')^{-1}. \tag{10}$$

Combining (6) and (10), we obtain the relativistic equation

$$1 - V^2/c^2 = (1 + V/u_x')(1 - V/u_x). \tag{11}$$

Solved for $u_x$, (11) leads to

$$u_x = \frac{u_x' + V}{1 + \frac{u_x' V}{c^2}} \tag{12}$$

whereas solved for $u_x'$, it leads to

$$u_x' = \frac{u_x - V}{1 - \frac{u_x V}{c^2}}. \tag{13}$$

From a single scenario and without intermediary steps, we have derived a relationship between the OX(O'X') components of the same bullet that moves with speed $u_x$ relative to K and with speed $u_x'$ relative to K', where K' moves with speed V relative to K and all the velocities are oriented in the positive direction of the common OX(O'X') axes.

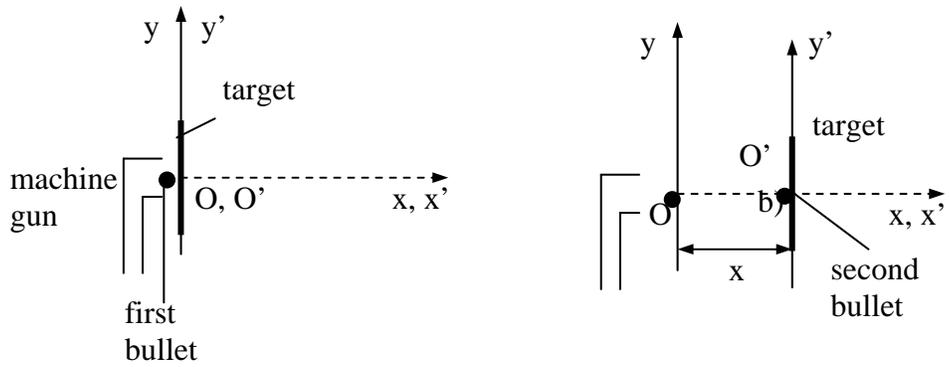

Figure 1. Scenario for deriving the addition law of relativistic velocities from time dilation.